\documentstyle[12pt,epsfig]{article}
\def\la{\mathrel{\mathpalette\fun <}}
\def\ga{\mathrel{\mathpalette\fun >}}
\def\fun#1#2{\lower3.6pt\vbox{\baselineskip0pt\lineskip.9pt
\ialign{$\mathsurround=0pt#1\hfil##\hfil$\crcr#2\crcr\sim\crcr}}}

\begin{document}

\begin{titlepage}
\vspace*{0.5cm}
\hspace{10cm} CMS NOTE 1998/085
 
\vspace{0.5cm}
\begin{center}
{\Large \bf 
Constraining the minimal supergravity model parameter $\tan\beta$
by measuring the dilepton mass distribution at LHC.}
\end{center}
 
\vspace{1.5cm}

\begin{center}
D.~Denegri$^1$, W.~Majerotto$^2$, and L.~Rurua$^{2,3}$

\vspace{1cm}
{\it $^1$ Centre d'Etudes Nucl$\acute{e}$aire de Saclay,
Gif-sur-Yvette,
France}

\vspace{0.3cm}
 {\it $^2$ Institut f$\ddot{u}$r Hochenergiephysik,
$\ddot{O}$sterreichische Akademie d.Wissenschaften, Vienna, Austria}

\vspace{0.3cm}
  {\it $^3$ Institute of Physics, Tbilisi, Georgia}
\end{center}

\vspace{2cm}

\begin{abstract}
We study the dependence on $\tan\beta$ of the event kinematics 
of final states with $e^+e^-\,/\mu^+\mu^-\,/e^{\pm}\mu^{\mp} \, +
E_T^{\rm miss} + jets$,
as expected in $pp$ collisions at CERN LHC, within the framework of the
minimal supergravity model.
With increase of $\tan\beta$, the third generation sparticle
masses $m_{\tilde{\tau}_1}$ and $m_{\tilde{b}_1}$ decrease
due to the increase of the tau and bottom Yukawa couplings. 
As a result, gluino, stop, sbottom, 
chargino and neutralino decays to third 
generation particles and sparticles are enhanced. With $\tan\beta$
rising, we
observe a characteristic change in the shape of the dilepton mass spectra
in $e^+e^-\,/\mu^+\mu^-\,+E_T^{\rm miss} + jets$ versus
$e^{\pm}\mu^{\mp} \, + E_T^{\rm miss} + jets$ final states, 
reflecting the presence of the decays 
$\tilde{\chi}_2^0\rightarrow \tilde{\ell}_{L,R}^{\pm}\ell^{\mp}\rightarrow
\tilde{\chi}_1^0 \ell^+\ell^-,\, \tilde{\chi}_2^0\rightarrow \tilde{\chi}_1^0
 \ell^+\ell^-$ and
$\tilde{\chi}_2^0\rightarrow \tilde{\tau}^{\pm}_1\tau^{\mp}\rightarrow
\tilde{\chi}_1^0 \tau^+ \tau^-$, 
$\tilde{\chi}_2^0\rightarrow \tau^+ \tau^- \tilde{\chi}_1^0$, respectively.
We exploit this effect for constraining the value of $\tan\beta$. 
\end{abstract}
\end{titlepage}

\section{Introduction}\label{sec1}
There is general agreement that new particles predicted by supersymmetry (SUSY)
have to be found at LHC, if the appealing conjecture
of low energy supersymmetry is correct. Therefore,
the discovery potential of LHC for SUSY particles has been studied 
extensively \cite{Bartl}.

Within the last years it became customary to study production and decay of SUSY
particles in the context of the 'Minimal Supergravity Model' (mSUGRA) 
\cite{Dine}. This model has only five parameters allowing one to study
systematically the whole parameter space. The mSUGRA model is based on the 
minimal supersymmetric extension of the Standard Model (MSSM), which has
the minimal possible content with two Higgs doublets (five physical Higgs
bosons). The number of parameters of the MSSM is reduced by unification 
conditions at the GUT point and by making use of the renormalisation group 
equation (RGE) to calculate the parameters at the electro-weak scale. 
A further strong constraint of the parameters is given by the requirement of 
spontaneous electroweak symmetry breaking at this scale (radiative
electroweak symmetry breaking). The parameters which remain are: $m_0$, the
common scalar mass at $M_{GUT}=10^{16}$ GeV, $m_{1/2}$ the unifying gaugino
mass, $A_0$ the common  trilinear term at $M_{GUT}$, $\tan\beta=v_2/v_1$
(with $v_1(v_2)$ being the Higgs vacuum expectation value of the Higgs 
$H_1^0(H_2^0)$), and $sign(\mu)$, $\mu$ being the higgsino mass parameter. This
model is incorporated in the Monte Carlo generator ISAJET \cite{BaerPaige}, 
which is used in our analysis.

A systematic study of all possible signals at LHC as a function of the mSUGRA
parameter space was carried out in \cite{Chen,CMScollab}. 
In \cite{Majer} we worked out a method  to 
determine the model parameters $m_0$ and $m_{1/2}$ with fixed $\tan\beta$
from an analysis of the event kinematics of final states with 
two same--flavor opposite--sign+ $E_T^{miss} + (jets)$, arising from
the decays $\tilde{\chi}_2^0\rightarrow \tilde{\ell}_{L,R}^{\pm}\ell^{\mp}
\rightarrow \ell^+ \ell^- \tilde{\chi}_1^0$, $\tilde{\chi}_2^0\rightarrow
\ell^+ \ell^- \tilde{\chi}_1^0$, with $\ell=e,\mu$.
Both in \cite{Chen} and in \cite{Majer} 
the value of $\tan\beta$ was taken to be $\leq 10$.
As, however, pointed out in \cite{Baer1,Bartl2,Baer3} at large $\tan\beta$ 
($\tan\beta\ga10$) the tau and bottom Yukawa coupling 
$f_{b,\tau}=\frac{gm_{b,\tau}}{\sqrt{2}m_W\cos\beta}$ may become comparable to 
the electroweak gauge coupling and even to the top Yukawa coupling
$f_t=\frac{gm_t}{\sqrt{2}m_W\sin\beta}$. As a consequence,
gluino, stop, sbottom, chargino and
neutralino decays into third generation particles and sparticles 
are enhanced when $\tan\beta$ is large \cite{Baer3,Baer4}. 
These features make the expected experimental signatures 
very different from those at low or moderate $\tan\beta$. 

Exploiting the fact that decays into $\tau$'s are enhanced at large
$\tan\beta$, one can get information about the underlying model framework.
With this aim, in \cite{Baer3,Baer4} 
the reconstruction of $\tau$'s by their {\it hadronic}
decays has been considered. In this paper
we study the decays $\tilde{\chi}_2^0\rightarrow \tilde{\tau}_1^{\pm}\tau^{\mp}
\rightarrow \tilde{\chi}_1^0 \tau^+\tau^-$ and
$\tilde{\chi}_2^0\rightarrow \tilde{\chi}_1^0 \tau^+\tau^-$, 
where the $\tau$'s decay {\it leptonically}  leading to final states with two 
{\it different}--flavor opposite--sign leptons + $E_T^{miss}$ + jets
as a characteristic experimental signature.

In Sec. \ref{sec2} we discuss the sparticle mass spectrum within mSUGRA.
Section \ref{sec3} is dedicated to a discussion of the production and
leptonic decays of $\tilde{\chi}_2^0$ in the mSUGRA parameter space.
In Sec. \ref{sec4} we work out a method to constrain
the parameter $\tan\beta$ by analysing the shape of the invariant mass
distributions in two same- and different--flavor
opposite--sign leptons + $E_T^{miss}$ + jets channels. 
We summarise our results and give some conclusions
in Sec. \ref{sec5}.

\section{Sparticle masses in mSUGRA}\label{sec2}
The masses of the sfermions (squarks and sleptons) of the first two 
generations are given by the RG equations,  
\cite{Kakuto}:

\begin{equation} \label{eqn1}
m^{2}_{\tilde{\ell}_{R}} = m^{2}_{0} + 0.15 m^{2}_{1/2} -
\sin^{2}\theta_{W} D 
\end{equation}

\begin{equation} \label{eqn2}
m^{2}_{\tilde{\ell}_{L}} = m^{2}_{0} + 0.52 m^{2}_{1/2} - 1/2 (1 -
2\sin^{2}\theta_{W}) D
\end{equation}

\begin{equation} \label{eqn3}
m^{2}_{\tilde{\nu}} = m^{2}_{0} + 0.52 m^{2}_{1/2} + 1/2 D 
\end{equation}

\begin{equation} \label{eqn4}
m^{2}_{\tilde{u}_{R}} = m^{2}_{0} + (0.07+c) m^{2}_{1/2} + 
2/3\sin^{2}\theta_{W} D 
\end{equation}

\begin{equation} \label{eqn5}
m^{2}_{\tilde{d}_{R}} = m^{2}_{0} + (0.02+c) m^{2}_{1/2} - 
1/3\sin^{2}\theta_{W} D 
\end{equation}

\begin{equation} \label{eqn6}
m^{2}_{\tilde{u}_{L}} = m^{2}_{0} + (0.47+c) m^{2}_{1/2} + 
(1/2-2/3\sin^{2}\theta_{W}) D 
\end{equation}

\begin{equation} \label{eqn7}
m^{2}_{\tilde{d}_{L}} = m^{2}_{0} + (0.47+c) m^{2}_{1/2} - 
(1/2-1/3\sin^{2}\theta_{W}) D 
\end{equation}

where $D=M^{2}_{Z} \cos2\beta$ and $4.5<c<6$.
In the third generation of sfermions $\tilde{f}$ the $\tilde{f}_L-\tilde{f}_R$
mixing may play an important r\^{o}le. The mass matrix in the basis 
$(\tilde{f}_L,\tilde{f}_R)$ with $f=t,b,\tau$ is given by: 

\vspace{2mm}

\begin{equation}\label{eqn8}
  {\cal M}_{\tilde{f}}^2 =
  \left( 
\begin{array}{cc} 
m^2_{\tilde{f}_L} & a_f\, m_f         \\[2mm]
      a_f\, m_f   & m^2_{\tilde{f}_R} 
\end{array}
\right)
\end{equation}

with 
\begin{equation}\label{eqn9}
m^2_{\tilde{f}_L}=M^2_{\tilde{Q},\tilde{L}} + m^2_Z\cos2\beta(I^f_{3L}-
e_f\sin^2\theta_W) + m^2_f 
\end{equation}

\begin{equation}\label{eqn10}
m^2_{\tilde{f}_R}=M^2_{\{\tilde{U},\tilde{D},\tilde{R}\}} + 
m^2_Z \cos2\beta e_f\sin^2\theta_W + m^2_f
\end{equation}

\begin{eqnarray}\label{eqn11}
a_f \,m_f =
\left\{ 
\begin{array}{cc}
(A_f-\mu\cot\beta)m_f & $for\,\,$  \tilde{f}=\tilde{t}\\
(A_f-\mu\tan\beta)m_f & $\,\,\,\,for\,\,$  \tilde{f}=\tilde{b},\tilde{\tau}
\end{array}
\right. 
\end{eqnarray}
where 
 $I_3^f$ is the third component of the weak isospin and $e_f$ the 
electric charge of the fermion $f$. $M_{\tilde{Q},\tilde{L},\tilde{U},
\tilde{D},\tilde{R}}$ and $A_f$ are soft SUSY-breaking parameters.
Note that the mixing term (11) is proportional to the fermion mass and 
$\cot\beta$ enters in the case of the stop and $\tan\beta$ in that of the 
sbottom and the stau. Therefore, strong mixing is expected for the stops,
but mixing may also be important for sbottom and stau if $\tan\beta$ is large
($\tan\beta\geq 10$) so that one mass eigenstate can be rather light.
The mass eigenstates $\tilde{f}_1$ and $\tilde{f}_2$ with masses 
$m_{\tilde{f}_1}$ and $m_{\tilde{f}_2}$ are obtained by diagonalizing the
matrix (8):

\begin{equation}\label{eqn12}
m^2_{\tilde{f}_1,\tilde{f}_2}=1/2(m^2_{\tilde{f}_L}+ 
m^2_{\tilde{f}_R})\mp 1/2 
\sqrt{(m^2_{\tilde{f}_L}-m^2_{\tilde{f}_R})+4a_f m_f}.
\end{equation}

In addition, one expects from the 
RG equations that due to the Yukawa interactions
the soft SUSY breaking masses $M_{\tilde{Q},\tilde{L},\tilde{U},
\tilde{D},\tilde{R}}$  of the 3rd generation sfermions are smaller 
than those  
of the 1st and 2nd generation. As in mSUGRA $|\mu|$ is quite generally 
large ($|\mu|>M$) the parameter $A_f$ does not play a crucial r\^{o}le,
especially for large $\tan\beta$ in the case of $\tilde{b}$ and $\tilde{\tau}$.
We therefore have taken $A=0$ in this study. 

Within the MSSM the masses of the charginos and neutralinos
are determined by the parameters $M=m_{1/2}(M_Z)$, $\mu$, $\tan\beta$
using 

\begin{equation}\label{eqn13}
M_1\simeq \frac{5}{3} \tan^2\theta_W M \simeq 0.5M, 
\end{equation}

$M_1$ being the $U(1)$
gaugino mass. In mSUGRA one has quite generally the following mass spectrum:

\begin{equation}\label{eqn14}
m_{\tilde{\chi}_2^0}\simeq m_{\tilde{\chi}_1^+}\simeq 2m_{\tilde{\chi}_1^0}
\simeq M=0.8 m_{1/2},  
\,\,m_{\tilde{\chi}_3^0}\simeq m_{\tilde{\chi}_4^0} \simeq |\mu|.
\end{equation}

In this model $\tilde{\chi}_1^0$ is almost  a pure $B-ino$ and
$\tilde{\chi}_2^0$  almost a pure $W^3-ino$. 
$\tilde{\chi}_1^+$ also has a strong gaugino component. On the other hand,
the heavier states $\tilde{\chi}_3^0$, $\tilde{\chi}_4^0$,
$\tilde{\chi}_2^{\pm}$ are higgsino-like. The gluino mass is given by

\begin{equation}\label{eqn15}
m_{\tilde{g}}=\frac{\alpha_s}{\alpha_2} M \simeq 3M \simeq 2.4 m_{1/2}
\end{equation}

\section{Production and leptonic decay of $\tilde{\chi}_2^0$ at large 
$\tan\beta$}\label{sec3}
At a hadron collider  
the neutralinos $\tilde{\chi}_2^0$ are dominantly produced in the decay chain
of massive gluinos and squarks, for instance $\tilde{g}\rightarrow q \bar q 
\tilde{\chi}_i^0$ or $\tilde{q}_{1,2}\rightarrow q \tilde{\chi}_i^0$,
$i=1,4$,
and $\tilde{\chi}_j^0\rightarrow \tilde{\chi}_k^0 Z^0$, $j=2,4$, $k=1,j-1$. 
These decays have been studied extensively for 
$\tan\beta\la 10$ in \cite{Barger}. Squarks and gluino decays for 
$\tan\beta\ga 10$ have been discussed first in \cite{Bartl2}. The collider 
phenomenology with large $\tan\beta$ has been quite generally discussed in
\cite{Baer1,Baer3}, including the chargino and neutralino decays. 
There are two 
features which play an important r\^{o}le at large $\tan\beta$. First, as
shown in section \ref{sec2} the sbottom $\tilde{b}_1$ and stau $\tilde{\tau}_1$
become lighter. Second, the Yukawa couplings of b and $\tau$
$f_{b,\tau}=\frac{gm_{b,\tau}}{\sqrt{2}m_W\cos\beta}$ increase with
$\tan\beta$. Both effects lead to a significant enhancement of sbottom and
stop production and of decays of 
gluinos, stops, sbottoms, charginos and neutralinos into third generation
particles and sparticles. In particular, the branching ratios of the
$\tilde{\chi}_2^0$ to (s)tau's
are increasing for large $\tan\beta$ with respect to the corresponding decays
into (s)electrons and (s)muons.
In Fig. 1 we show the maximum possible branching ratio values 
(for $m_0\la 500$ GeV,  $m_{1/2}\la 900$ GeV, $\mu<0$) 
of the following leptonic decays of $\tilde{\chi}_2^0$, 
 as a function of
$\tan\beta$:

\begin{equation}\label{eqn16}
\tilde{\chi}_2^0 \rightarrow \tilde{\chi}_1^0 \ell^+ \ell^-, 
\end{equation}

\begin{equation}\label{eqn17}
\tilde{\chi}_2^0 \rightarrow \tilde{\ell}_L \ell, 
\end{equation}

\begin{equation}\label{eqn18}
\tilde{\chi}_2^0 \rightarrow \tilde{\ell}_R \ell,
\end{equation}

where $\ell=e$ and $\mu$, and 

\begin{equation}\label{eqn19}
\tilde{\chi}_2^0 \rightarrow \tilde{\tau}_1 \tau,
\end{equation}

\begin{equation}\label{eqn20}
\tilde{\chi}_2^0\rightarrow \tau^+ \tau^- \tilde{\chi}_1^0.
\end{equation}

In mSUGRA when $\tilde{\chi}_2^0$ decays into sleptons
are kinematically allowed,
the sleptons then decay directly into the LSP with 
$B(\tilde{\ell}^{\pm}_{L,R} \rightarrow \tilde{\chi}^{0}_{1}
\ell^{\pm})=100$\% and
$B(\tilde{\tau}_1\rightarrow \tau  \tilde{\chi}_1^0)\sim 100\%$.

As can be seen in Fig. 1, the branching ratio of $\tilde{\chi}_2^0
\rightarrow \tilde{\tau}_1\tau$ can even become $100\%$ for $\tan\beta\geq10$,
if this channel is the only two-body decay channel kinematically allowed.
Furthermore, the branching ratio of the three-body decay $\tilde{\chi}_2^0
\rightarrow \tilde{\chi}_1^0 \tau^+ \tau^-$ can go up to $90\%$. It does not
reach $100\%$ because there are always competing  other three-particle
decay channels, in particular $\tilde{\chi}_2^0
\rightarrow \tilde{\chi}_1^0 b \bar b$. 
The $\tau$'s decay as $\tau^-\rightarrow e^- \bar \nu_e \nu_{\tau}, 
\mu^- \bar \nu_{\mu} \nu_{\tau}$ with a branching ratio of $17.8\%$ and 
$17.4\%$, respectively. Therefore, with $\tan\beta\geq10$ the decays
$\tilde{\chi}_2^0\rightarrow \tilde{\tau}_1 \tau \rightarrow
\tau^+ \tau^- \tilde{\chi}_1^0$ and  $\tilde{\chi}_2^0\rightarrow  
\tau^+ \tau^- \tilde{\chi}_1^0$ give a 
contribution to final states with two leptons with opposite charge and the
{\it same flavor}, whereas with $\tan\beta\la10$ that contribution is 
negligible
compared to that from $\tilde{\chi}_2^0\rightarrow \tilde{\ell}_{L,R}^{\pm}
\ell^{\mp} \rightarrow \ell^+\ell^-\tilde{\chi}_1^0$,  
$\tilde{\chi}_2^0\rightarrow \ell^+\ell^-\tilde{\chi}_1^0$ with $\ell=e,\mu$.
However, the distinctive feature of the $\tilde{\chi}_2^0$ decays into 
$\tau$'s is that they also lead to final states with two opposite sign but 
{\it different flavor}  leptons.

In Fig. 2 we show for $\tan\beta=10$, $\mu>0$ in the $(m_0,m_{1/2})$
plane contour lines for 
cross-section times branching ratios ($\sigma \times B$)
for production of $\tilde{\chi}_2^0$ followed by decays
$\tilde{\chi}_2^0 \rightarrow \tilde{\ell}_{L,R}^{\pm}\ell^{\mp}\rightarrow 
\ell^+ \ell^- \tilde{\chi}_1^0\,$, 
$\tilde{\chi}_2^0 \rightarrow \ell^+ \ell^- \tilde{\chi}_1^0\,$, 
$\tilde{\chi}_2^0 \rightarrow \tilde{\tau}_1^{\pm}
\tau^{\mp}\rightarrow \tau^+ \tau^- \tilde{\chi}_1^0
\rightarrow \ell^+ \nu_{\ell} \bar \nu_{\tau}\, \ell^-
\bar \nu_{\ell} \nu_{\tau} \tilde{\chi}_1^0\,$,  
$\tilde{\chi}_2^0 \rightarrow \tau^+ \tau^- \tilde{\chi}_1^0 \rightarrow 
\ell^+ \nu_{\ell} \bar \nu_{\tau} \,\ell^-
\bar \nu_{\ell} \nu_{\tau} \tilde{\chi}_1^0$,  
with $\ell=e$ and $\mu$.
As in the low $\tan\beta$ case we can again define three kinematical
domains, determined by the masses of $\tilde{\chi}_2^0$ and the sleptons
\cite{Majer}, with the characteristic decays:

\begin{tabular}{lrcl}
domain I ($m_{0} \ga 0.6 m_{1/2}, m_{1/2} \la 250$~GeV): &
$m_{\tilde{\chi}^{0}_{2}} \la m_{\tilde{\tau}_1}$ & , & with
$\tilde{\chi}^{0}_{2} \rightarrow \tilde{\chi}^{0}_{1}\ell\ell$ \\
& & & and $\tilde{\chi}^{0}_{2} \rightarrow \tilde{\chi}_1^0 \tau \tau$\\
\\
domain II ($0.3 m_{1/2} \la m_{0} \la 0.6  m_{1/2}$):
& $m_{\tilde{\ell}_{R}} < m_{\tilde{\chi}^{0}_{2}} <
m_{\tilde{\ell}_{L}}$ & , & with $\tilde{\chi}^{0}_{2} \rightarrow
\tilde{\ell}_{R} \ell$ \\
 & & & and $\tilde{\chi}^{0}_{2} \rightarrow \tilde{\tau}_1 \tau$ \\
\\
domain III ($m_{0} \la 0.3  m_{1/2} )$: &
$m_{\tilde{\chi}^{0}_{2}} \ga m_{\tilde{\ell}_{L}}$ & , & with
$\tilde{\chi}^{0}_{2} \rightarrow \tilde{\ell}_{L} \ell$
\end{tabular}

\bigskip

In domains I and II there is an additional source of two opposite--sign
lepton final states from decays to tau's and stau's respectively,
if $\tan\beta\ga 10$. 
In domain III the decays $\tilde{\chi}^{0}_{2} \rightarrow 
\tilde{\ell}_{R} \ell$ and $\tilde{\chi}^{0}_{2} \rightarrow \tilde{\tau}_1 
\tau$ would be kinematically allowed. As $\tilde{\chi}^{0}_{2}$ is almost
a $W^3$-ino these decays are suppressed because $\tilde{\tau}_1 \simeq 
\tilde{\tau}_R$.   

Moreover, comparing to the $\tan\beta=2$ case \cite{Majer}, 
one can observe a significant decrease of $\sigma \times B$ values for 
$\tilde{\chi}_2^0\rightarrow \tilde{\ell}_{L,R}\ell\rightarrow
 \ell^+ \ell^- \tilde{\chi}_1^0$
and $\tilde{\chi}_2^0\rightarrow \ell^+ \ell^- \tilde{\chi}_1^0$ decays. 
For $\tan\beta \ga 25$ the domain for $\tilde{\chi}_2^0\rightarrow
\tilde{\ell}_{L,R}\ell$ almost disappears due to the dominance
of the decays  $\tilde{\chi}_2^0\rightarrow \tilde{\tau}_1\tau$, see Fig. 1.
This can be seen in Fig. 3 where we show the same as in Fig. 2,
but for $\tan\beta=35$. Here we have only two domains left which are:

\begin{tabular}{lrcl}
domain I ($m_{0} \ga m_{1/2}, m_{1/2} \la 250$~GeV): &
$m_{\tilde{\chi}^{0}_{2}} \la m_{\tilde{\tau}_1}$ & , & with
$\tilde{\chi}^{0}_{2} \rightarrow \tilde{\chi}^{0}_{1}\ell\ell$ \\
& & & and $\tilde{\chi}^{0}_{2} \rightarrow \tilde{\chi}_1^0 \tau \tau$\\
\\
domain II ($0.5 m_{1/2} \la m_{0} \la  m_{1/2}$):
& $m_{\tilde{\tau}_1} < m_{\tilde{\chi}^{0}_{2}} <
m_{\tilde{\ell}_{R}}$ & , & with 
$\tilde{\chi}^{0}_{2} \rightarrow \tilde{\tau}_1 \tau$ 
\end{tabular}

\bigskip
Moreover, as one can see comparing Figs. 2 and 3, 
the upper theoretically excluded area 
is increasing with $\tan\beta$. This is due to the fact that the stau lepton
mass becomes lighter than that of $\tilde{\chi}_1^0$ with values
of $m_0$ and $m_{1/2}$ decreasing.

Another interesting feature appearing at large $\tan\beta$ is the enhanced 
yield of $Z^0$. This is due to three factors: the larger production cross 
section for $\tilde{b}_1\tilde{b}_1$ because of the smaller $\tilde{b}_1$
mass, the enhanced branching ratio of $\tilde{b}_1\rightarrow 
\tilde{\chi}^0_3 b$ due to the Yukawa coupling of b to the
Higgsino component of
$\tilde{\chi}_3^0$, and to the decay $\tilde{\chi}_3^0\rightarrow Z^0
\tilde{\chi}_{1,2}^0$. 

Furthermore, in the region $m_{1/2}\ga 250$ GeV, where the decay 
$\tilde{\chi}_2^0\rightarrow Z^0
\tilde{\chi}_1^0$ is kinematically possible, it is generally enhanced compared 
to low $\tan\beta$, because of the larger Higgsino component of 
$\tilde{\chi}_2^0$ and $\tilde{\chi}_1^0$ at large $\tan\beta$. There is
also a narrow band 260 GeV$\la m_{1/2} \la 305$ GeV, where the decay 
$\tilde{\chi}_2^0\rightarrow Z^0 \tilde{\chi}_1^0$ is the only 
two-particle decay kinematically possible because the mass of $h^0$
is larger at large $\tan\beta$.

\section{Constraining $\tan\beta$}\label{sec4}
\subsection{Signatures}
At low $\tan\beta$, the decays of $\tilde{\chi}_2^0$ to (s)electrons and
(s)muons, Eqs. (16)-(18), dominate over  those into (s)tau's, Eqs. (19)-(20), 
leading to electron and muon pairs in the final states, see Fig. 1.
The favored signature to select the decay channels Eqs. (16)-(18) is given by
two {\it same--flavor} opposite-sign leptons $+ E_T^{miss}$ + jets final states
(SFOS). 
As is known from previous works \cite{Baer2,Paige,CMScollab}, 
the invariant mass 
distribution of the two leptons has a pronounced edge at the kinematical 
end--point.  
With increase of $\tan\beta$ the decays to third generation 
(s)particles  (Eqs. (19)-(20)) are increasing.
To select these decays we use the event topology of two {\it different--flavor} 
opposite--sign leptons $+ E_T^{miss}$ + jets (DFOS). 

\subsection{Signal and background event simulation}
The simulations are done at the particle level with parametrised detector
responses based on detailed detector simulations. These parametrisations
are adequate for the level of detector properties we want to
investigate, and are the only practical ones in view of the multiplicity
and complexity of the final state signal and background channels
studied. The essential ingredients for the investigation of SUSY
channels are the response to jets, $E_T^{\rm miss}$, the lepton identification
and isolation capabilities of the detector, and the capability to tag b-jets. 
The CMS detector simulation program CMSJET 3.2 \cite{Khanov} is used.
It contains all significant detector response aspects, calorimeter
acceptances and resolutions, granularity, main detector cracks and effects of 
magnetic field.
For more details we refer to Ref. \cite{Majer,CMScollab}.

Standard Model background processes are generated with PYTHIA 5.7
\cite{Sjostrand}. We use CTEQ2L structure functions.
The largest  background is due to $t\bar{t}$ production,
with both $W$'s decaying into leptons, or one of the leptons  from a $W$
decay and the other from the $b$-decay of the same $t$-quark.  We also
considered other Standard Model (SM)
backgrounds: $W+jets$, WW, WZ, $Z+jets$,  ZZ, $b \bar b$
and  $\tau \tau$-pair production, with decays into electrons and muons.
Chargino pair production $\tilde{\chi}_1^{\pm}\tilde{\chi}_1^{\mp}$ is
the largest SUSY background, but gives a small contribution compared
to the signal.

\subsection{The shape of the dilepton mass distribution in the mSUGRA parameter
space}
In this section we will analyse the shape of the invariant 
mass distribution of the two leptons $\ell=e,\mu$,
coming from the $\tilde{\chi}_2^0$ 
decays, Eqs. (\ref{eqn16})-(\ref{eqn20}), considering SFOS and DFOS channels.
To this purpose we have investigated the expected dilepton invariant mass 
distributions for different values of the parameters $m_0, \, m_{1/2}, \, 
\tan\beta$, with $\mu>0$. 

In Figs. 4(a)-4(d) we show the invariant mass 
distributions of the two leptons in the SFOS and DFOS channels
at a point 
$m_0=100$ GeV, $m_{1/2}=190$ GeV, for $\tan\beta=2,\, 10,\, 15,\,25$, 
respectively ($\mu>0$). 
The mass spectrum at this point is: $m_{\tilde{\chi}_2^0}\approx 140$ GeV,
$m_{\tilde{\chi}_1^0}\approx 75$ GeV, $m_{\tilde{\ell}_R}\approx 132$ GeV,
but $m_{\tilde{\tau}_1}$ decreases from 124 GeV to 91 GeV with the
increase of
$\tan\beta$.
 
At this ($m_0,m_{1/2}$) point, which belongs 
to domain II, the two-step decays $\tilde{\chi}_2^0 \rightarrow 
\tilde{\ell}_R^{\pm} \ell^{\mp}\rightarrow \ell^+ \ell^- \tilde{\chi}_1^0$ 
are possible for $\tan\beta\la 25$, but the decays  
$\tilde{\chi}_2^0\rightarrow \tilde{\tau}_1^{\pm}\tau^{\mp} \
\rightarrow \tau^+ \tau^- 
\tilde{\chi}_1^0$ start to contribute significantly
to the final states considered
for $\tan\beta\ga 10$.
Thus, in Figs. 4(a)-4(d)  
one can observe a characteristic change of the dilepton mass shape in  the
SFOS  and  DFOS channels with increase of $\tan\beta$,
reflecting the appearance or disappearance of the corresponding decay modes.

For low $\tan\beta$ ($\la 10$), a pronounced edge is visible in the
$M_{\ell^+\ell^-}$ distribution in the SFOS channel 
with the maximum at \cite{Paige}:
 
\begin{equation}\label{eqn21}
M^{max}_{\ell^+\ell^-} =  \frac{\sqrt{ (m^2_{\tilde{\chi}^0_2} -
m^2_{\tilde{\ell}_R})
(m^2_{\tilde{\ell}_R} - m^2_{\tilde{\chi}^0_1}) }} {m_{\tilde{\ell}_R} },
\end{equation}
while no characteristic structure in the $M_{\ell^+\ell^{'-}}$ spectrum
is expected in the DFOS case, see Fig. 4(a) for
$\tan\beta=2$. In the DFOS case the difference in magnitude (but not in shape)
between the 
$M_{\ell^+\ell^{'-}}$ spectrum predicted by the SM and that by mSUGRA 
is due to internal SUSY background, mainly
due to $\tilde{\chi}_1^+ \tilde{\chi}_1^-$ pair production from gluino/squark
cascade decays.

Fig. 4(b) is as Fig. 4(a), but for $\tan\beta=10$. 
In comparison to the $\tan\beta=2$
case one can see a decrease of the statistics in both the
SFOS and DFOS channels for the same integrated luminosity of $5\times10^3$ 
pb$^{-1}$. In both SFOS and DFOS final states,
the event rates are by an order 
of magnitude larger than those according to SM expectations.
Moreover, a pronounced deviation in the shape of the  $M_{\ell^+\ell^{'-}}$
spectrum from the expected SM and SUSY background is observable 
in the low mass region. 
The corresponding $M_{\tau^+\tau^-}$ spectrum due to the
$\tilde{\chi}_2^0$ decays to stau's should have a maximum at: 
\begin{equation}\label{eqn22}
M^{max}_{\tau^+\tau^-} 
=  \frac{\sqrt{ (m^2_{\tilde{\chi}^0_2} - m^2_{\tilde{\tau}_1})
(m^2_{\tilde{\tau}_1} - m^2_{\tilde{\chi}^0_1}) }} {m_{\tilde{\tau}_1} }.
\end{equation}

The edge of the ditau spectrum is shifted 
from $M_{\ell^+\ell^-}^{max}=37$ GeV to
$M^{max}_{\tau^+\tau^-}=51$ GeV due to the 
mass difference between $\tilde{\ell}_R$ and $\tilde{\tau}_1$, 
see Eqs. (21) and (22). But the spectrum of the DFOS channel 
proceeding through $\tau\rightarrow\ell\nu\nu$ decays 
is not so pronounced  
as in the SFOS channel having no sharp edge due to the missing momentum
taken by four neutrinos from $\tau$ decays.
Therefore the $\ell^+\ell^{'-}$ mass is distributed in 
the lower mass region, below the expected ditau kinematical end point.
Thus, a distinctive feature of the $M_{\ell^+\ell^{'-}}$ spectrum is  
an enhancement in both shape and magnitude over background
in the low mass range ("low--mass enhancement").

A further characteristic difference in the dilepton mass spectrum 
compared to $\tan\beta=2$ is the appearance of a $Z^0$ peak in the SFOS
channel, which  can clearly be seen in Fig. 4(b). Here the $Z^0$ peak is 
due to the decays $\tilde{\chi}_3^0\rightarrow Z^0\tilde{\chi}_2^0$, 
as explained in Sec. \ref{sec3}. 

With further increase of $\tan\beta$, due to the significant increase
of $\tilde{\chi}_2^0$ decays into stau's and the corresponding decrease of 
decays into selectrons and smuons, we observe a deterioration of the
sharpness of the edge also in the SFOS channel, 
see Fig. 4(c) for $\tan\beta=15$.
Another consequence of the change in the relative branching fractions of
these decays is that the event rate difference between 
the SFOS and DFOS channels  
decreases. This effect is visible by comparing  Figs. 4(b) and 4(c).
Figure 4(c) also exhibits a decrease of
the relative event rate below the two peaks from $\tilde{\chi}_2^0
\rightarrow \tilde{\ell}\ell/\tilde{\tau}_1\tau$ and 
$\tilde{\chi}_3^0\rightarrow Z^0\tilde{\chi}_2^0$ decays. 

Finally, at $\tan\beta\simeq25$, Fig. 4(d), when the decays into 
stau's are the only decay modes contributing to the final states considered,
the dilepton mass spectra in the SFOS and DFOS channels become similar with
complete disappearance of sharp edges but a strong low--mass enhancement,
and the corresponding event rates become comparable.
The low--mass $\ell^+\ell^-$ peak and the $Z^0$ peak are of the same magnitude.

For comparison of the dilepton mass spectra at this 
($m_0,\,m_{1/2}$) point we used a common set of cuts for all values of 
$\tan\beta$, requiring two hard isolated leptons
with $p_T^{\ell_{1,2}}\ge 10$ GeV in $|\eta|<2.4$,
accompanied by large transverse
missing energy, $E_T^{\rm miss}\ge 150$ GeV and jet multiplicity $N_{jet}\ge 3$
with energy $E_T^{jet}\ge 60$ GeV in the rapidity range
$|\eta_{jet}|<4.5$. This point is reachable at low luminosity
${\cal L}_{int}=5\times 10^3$ pb$^{-1}$ even at the largest $\tan\beta$ shown.
The observability of the edges in both the 
SFOS and DFOS channels is estimated by the criterion 
$(N_{EV}-N_B)/\sqrt{N_{EV}}\ga 5$ and $(N_{EV}-N_B)/N_B \ga 1.3$, where
$N_{EV}$ is the number of events with $M_{\ell^+\ell^{(')-}} \le
M_{\ell^+\ell^{(')-}}^{max}$,
and $N_B$ is number of the expected SM and SUSY 
background events in the same mass range.
In case the edge is not very sharp due to the
dominance of $\tilde{\chi}_2^0$ decays to stau's,  
we estimate in practice the kinematical "end-points"
$M_{\ell^+\ell^-}^{max}$ and $M_{\ell^+\ell^{'-}}^{max}$
by subtracting the event sample with the same kinematics, but 
containing two same-sign leptons for the DFOS channel, 
or different-flavor events for the SFOS channel, 
normalized to the non-threshold region.
The precision of the kinematical end-point measurement in this case is 
estimated to be $\pm 5$ GeV for $M_{\ell^+\ell^{(')-}}^{max}\la 50$ GeV.  
For  $M_{\ell^+\ell^{(')-}}^{max}\ga 50$ GeV the mass difference
between $\tilde{\chi}_2^0$ and the sleptons is larger, and the
neutrinos can have larger momentum. 
In this case the technique of reconstructing the
kinematical end-point requires a more detailed study
and is beyond the scope of this paper.

To illustrate the behaviour of the
$M_{\ell^+\ell^-}$ and $M_{\ell^+\ell^{'-}}$
spectra at very large $\tan\beta$,
we show in Fig. 5 various ($m_0,\,m_{1/2}$) points with $\tan\beta=35$.
Figure 5(a) clearly
exhibits the similarity of the dilepton mass spectra in the SFOS 
and DFOS channels with comparable event rates,
as expected at large $\tan\beta$ from $\tilde{\chi}_2^0$
two--body decays to stau's (domain II).
Figures 5(b) represents a ($m_0,\,m_{1/2}$) point where three-body decays
of $\tilde{\chi}_2^0$ are possible. 
Here the two contributions from the decays 
Eqs. (16) and (20) are visible in the SFOS channel, while in the DFOS channel
only $\tilde{\chi}_2^0$ decays into tau's are selected, see Fig. 5(b). 
Figure 5(c) shows the $M_{\ell^+\ell^-}$
spectrum at point $m_0=800$ GeV, $m_{1/2}=250$ GeV, $\tan\beta=35$, $\mu>0$
with the decays $\tilde{\chi}_2^0\rightarrow \ell^+\ell^-
\tilde{\chi}_1^0$, 
$\ell=e,\mu$. In Figs. 5(a)-5(c) the $Z^0$ peak is higher than  the 
"low--mass enhancement" as being expected for large $\tan\beta\ga 25$.
For $m_{1/2}\ga 450$ GeV, Fig. 5(d), there is only a 
$Z^0$ peak left, mainly due to the decays
$\tilde{\chi}_3^0\rightarrow Z^0\tilde{\chi}_{1,2}^0$.

To reach  the ($m_0,\,m_{1/2}$) points shown in Fig. 5, 
a higher luminosity is necessary,
${\cal L}_{int}=5\times10^4-10^5$ pb$^{-1}$. Correspondingly harder cuts
on the leptons and the missing energy have to be applied: 
two isolated leptons with $p_T^{\ell_{1,2}}\ge 10-15$ GeV, 
$E_T^{\rm miss}\ge 200-300$ GeV, $N_{jet}\ge 3$
with $E_T^{jet}\ge 60$ GeV in  $|\eta_{jet}|<4.5$. 
The parameter region reachable at $\tan\beta=35$ 
is limited with $\sigma\times B\sim0.01$ pb for $\tilde{\chi}_2^0$ decays
into stau's and $\sigma\times B\sim0.1$ pb for $\tilde{\chi}_2^0
\rightarrow \ell^+ \ell^- \tilde{\chi}_1^0$, see Fig. 3. 
 
\section{Summary and Conclusions}\label{sec5}
In this paper we worked out a method to constrain the parameter
$\tan\beta$ of the Minimal Supergravity Model in pp collisions at LHC. 
The characteristic features arising with increase of $\tan\beta$ are:

1) the $\tilde{\chi}_2^0$ decay branching ratio into (s)taus 
is increasing, whereas the decay fraction into (s)electrons and (s)muons is 
decreasing;
 
2) $\tilde{\chi}_3^0$ production and decays into $Z^0$ are enhanced.\\
\\
The method is based on the analysis of the dilepton mass distributions  
in the 
simplest experimental signatures, $e^+e^-/\mu^+\mu^- + E_T^{\rm miss} + jets$ 
and $e^{\pm}\mu^{\mp} + E_T^{\rm miss} + jets$. These signatures
select different leptonic decays of the neutralinos and hence are  
sensitive to the value of $\tan\beta$.
In particular,
events with $e^+e^-/\mu^+\mu^- + E_T^{\rm miss} + jets$ final states
mainly contain the decay products from  $\tilde{\chi}_2^0$ decays to  
(s)electrons and (s)muons at low $\tan\beta$ and to leptonically
decaying stau's at large $\tan\beta$.
Moreover, at large $\tan\beta$, the channel 
$e^+e^-/\mu^+\mu^- + E_T^{\rm miss} + jets$ allows one to measure the
$Z^0$ boson. At large $\tan\beta$ the topology with
$e^{\pm}\mu^{\mp} + E_T^{\rm miss} + jets$ selects
tau's decaying leptonically.

The dilepton mass distributions from the decays of $\tilde{\chi}_2^0$ to 
(s)electrons/(s)muons and to (s)tau's have different shapes. The decays
to (s)electrons and (s)muons lead to a pronounced edge at the kinematical 
end point in the $M_{\ell^+\ell^-}$ spectrum, while the edge due to 
the decays into (s)tau's with tau's decaying into electrons and 
muons is smeared out because of missing  
momentum taken away by four neutrinos. The dilepton mass spectrum, however,
has a pronounced low--mass enhancement, a peak below 
the kinematical end point. 
Therefore, the observation of a 
"low--mass enhancement" in shape and magnitude in the 
dilepton mass distribution is regarded as evidence for stau lepton production,
particularly so in different flavor final states.

We have made use of the expected characteristic dilepton mass shapes and event
rates for identifying the corresponding decays and thus for constraining the 
value of $\tan\beta$. 
An important aspect of the method is that no particularly
stringent demands 
on detector performance are required for identifying the tau's from the 
decays of neutralinos even at high luminosity. Essential is the ability to 
define and select isolated electrons and muons.
Another important aspect of this method is that it may give information
on $\tan\beta$ even with low integrated luminosity.

We have shown that:

$\bullet$ An edge in the $M_{\ell^+\ell^-}$ distribution is expected to be 
visible in the  $e^+e^-/\mu^+\mu^- + E_T^{\rm miss} + jets$ channel 
for $\tan\beta\la 25$ in domain II ($0.3m_{1/2}\la m_0\la0.6m_{1/2}$),
and for  $\tan\beta\la 50$ in domain I ($m_0\ga 0.6 m_{1/2},\, 
m_{1/2} \la 250$ GeV).
The method to differentiate between
two- and three-body  decays of $\tilde{\chi}_2^0$
was discussed in Ref. \cite{Majer};

$\bullet$ In the dilepton mass spectrum 
a clear deviation in shape and magnitude
from the SM and internal SUSY background is expected 
in $e^{\pm}\mu^{\mp} + E_T^{\rm miss} + jets$ final states with 
$\tan\beta \ga 10$;

$\bullet$ Two components in the dilepton mass spectrum 
due to three-body  $\tilde{\chi}_2^0$ decays to electrons/muons
and to tau's are observed for $\tan\beta \ga 10$.

Combining the results of the analysis of the dilepton mass distributions 
in $e^+e^-/\mu^+\mu^- + E_T^{\rm miss} + jets$ versus
$e^{\pm}\mu^{\mp} + E_T^{\rm miss} + jets$ final states, 
we have found the following criteria for constraining $\tan\beta$ 
in domain II:

$\bullet$ Observation of an edge in the dilepton mass spectrum in 
$e^+e^-\& \mu^+\mu^- + E_T^{\rm miss} + jets$  
channels, but no deviation from SM and internal SUSY background in that for
$e^{\pm}\mu^{\mp} + E_T^{\rm miss} + jets$ 
channel restricts  $\tan\beta$ to be $\la 10$;
 
$\bullet$ Observation of an edge in the dilepton mass spectra in 
$e^+e^-\& \mu^+\mu^- + E_T^{\rm miss} + jets$ and a low--mass enhancement in
$e^{\pm}\mu^{\mp} + E_T^{\rm miss} + jets$ channels
constrains the value of $\tan\beta$  to $10 \la \tan\beta\la 25$;

$\bullet$ Observation of the similar dilepton mass spectra  
in $e^+e^-\& \mu^+\mu^- + E_T^{\rm miss} + jets$ and 
$e^{\pm}\mu^{\mp} + E_T^{\rm miss} + jets$ channels 
with comparable shapes and event rates, indicates $\tan\beta\ga 25$.

A further constraint of $\tan\beta$  can be found by measuring the
$Z^0$ peak in the $e^+e^- \& \mu^+\mu^- + E_T^{\rm miss} + jets$
channel. This channel has the advantage of getting no contribution 
from SM background what makes the SUSY $Z^0$ signal 
event sample very clean. This is especially useful if an edge due to
$\tilde{\chi}_2^0$ three-body decays is observed.
The absence of a $Z^0$ peak will restrict $\tan\beta$ to be $\la2$. 
We have also analysed the relative event rate in the "low--mass peak" of the
$M_{\ell^+\ell^-}$ spectrum and in the $Z^0$ peak. 
If both peaks are of the same size the
value of $\tan\beta$ should be $\ga 25$.

In the region $10 \la \tan\beta\la 25$, $\tan\beta$ 
can be determined more precisely by measuring  the kinematical end points,
$M_{\ell^+\ell^-}^{max}$ in the $e^+e^-\&\mu^+\mu^-$ spectrum
and  $M_{\ell^+\ell^{'-}}^{max}$ in the $e^{\pm}\mu^{\mp}$ spectrum. 
Knowing the $M_{\ell^+\ell^-}^{max}$ contour line 
in the ($m_0,m_{1/2}$) parameter plane \cite{Majer} with good precision, we
can constrain $\tan\beta$ by using the relative event rates 
$\frac{N_1(e^+e^-\& \mu^+\mu^-)}{N_2(e^{\pm}\mu^{\mp})}$ and
$\frac{N_1(e^+e^-\&\mu^+\mu^-)}{N_3(Z^0)}$. Here  $N_1$ is the
number of events with $M_{\ell^+\ell^-}\la M_{\ell^+\ell^-}^{max}$,
$N_2$ is the number of events with 
$M_{\ell^+\ell^{'-}}\la M_{\ell^+\ell^{'-}}^{max}$,
and $N_3$ is the number of events in the mass region
of $Z^0$, 86 GeV $<m_{Z^0}<$ 96 GeV.
We have found that the $M_{\ell^+\ell^-}^{max}$ contour
lines in the ($m_0,m_{1/2}$) parameter plane are almost independent of 
$\tan\beta$ due to the fact that  the masses of
$\tilde{\chi}_2^0, \tilde{\chi}_1^0,
\tilde{\ell}_R$ do not strongly depend on $\tan\beta$ in mSUGRA.

Quite generally, the observation of a characteristic deviation in the 
dilepton mass shape 
from that predicted by the SM background 
can be considered an evidence for Physics Beyond the Standard Model,
in particular SUSY \cite{Majer}. 

\vspace{0.5cm}
{\Large \bf Acknowledgements}
 
L.R. thanks for financial support 
by the Austrian Academy of Sciences.
This work was also supported by the "Fonds zur F\"{o}rderung der
Wissenschaftlichen Forschung" of Austria, project no. P10843-PHY.

\clearpage
\begin{figure}[htbp]
\centering
\resizebox{120mm}{150mm}
{\includegraphics{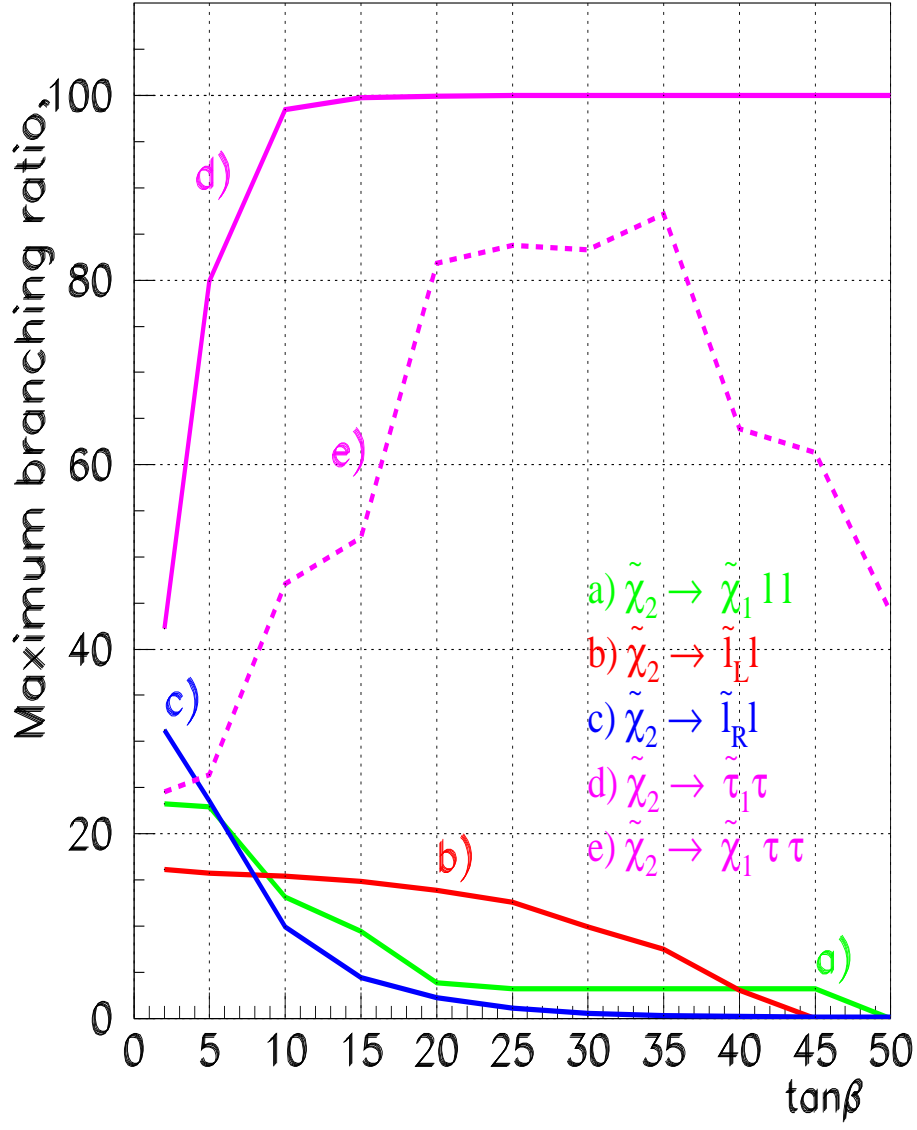}}
\caption{The maximum branching ratio of the decays:
a) $\tilde{\chi}_2^0 \rightarrow \tilde{\chi}_1^0 \ell^+ \ell^-$,
b) $\tilde{\chi}_2^0 \rightarrow  \tilde{\ell}_L^{\pm} \ell^{\mp}$,
c) $\tilde{\chi}_2^0 \rightarrow  \tilde{\ell}_R^{\pm} \ell^{\mp}$,
d) $\tilde{\chi}_2^0 \rightarrow \tilde{\tau}_1^{\pm}
   \tau^{\mp}$, and  
e) $\tilde{\chi}_2^0 \rightarrow \tau^+ \tau^- \tilde{\chi}_1^0$
with $\ell=e$ and $\mu$, for
$A_0=0, \, \mu > 0$.}
\end{figure}

\clearpage
\begin{figure}[htbp]
\centering
\resizebox{160mm}{180mm}
{\includegraphics{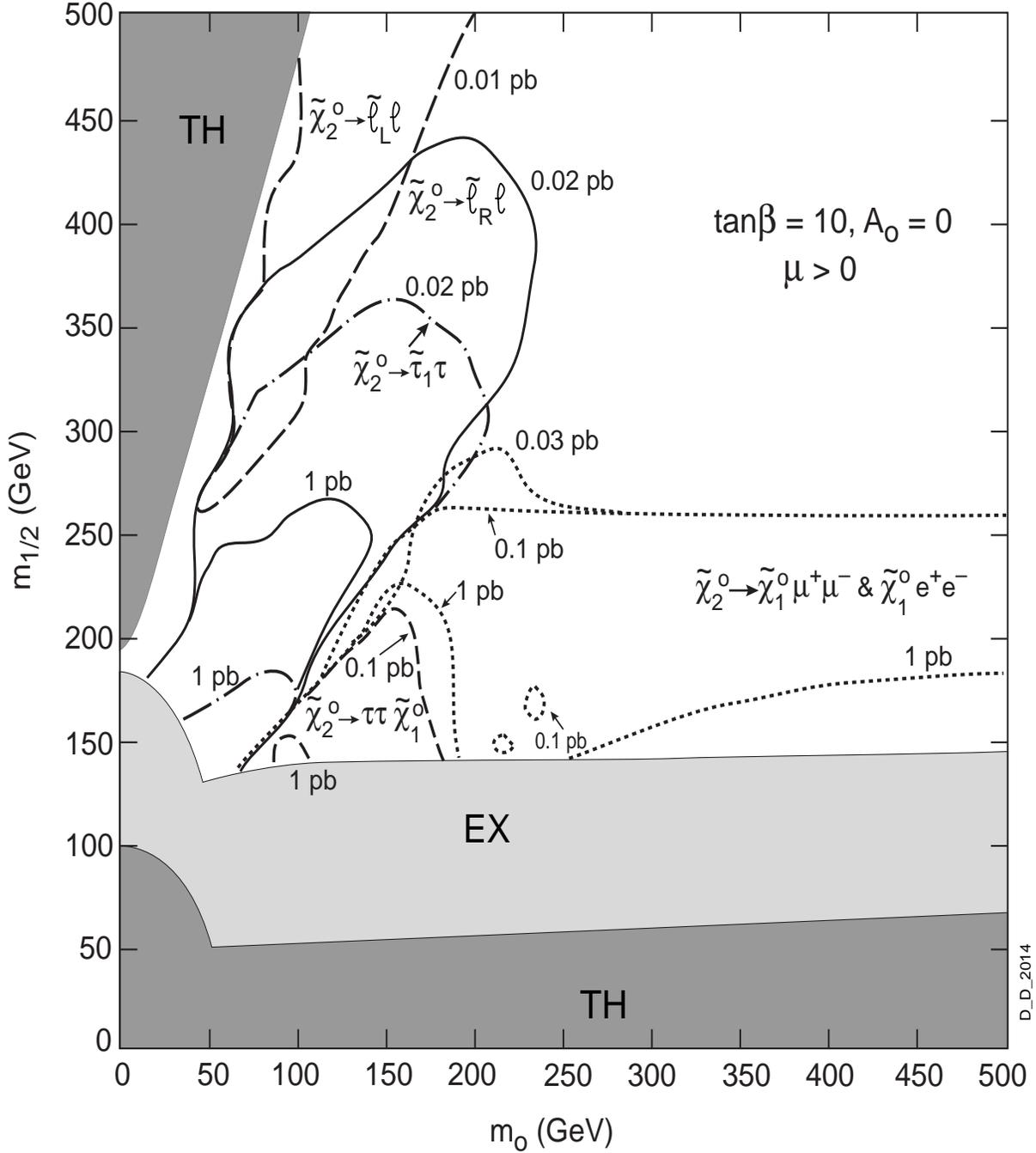}}
\caption{Contour lines for cross-section times branching ratios in the 
($m_0,m_{1/2}$) plane for indirect and associated $\tilde{\chi}_2^0$ production
followed by decays:
$\tilde{\chi}_2^0 \rightarrow  \tilde{\ell}_R^{\pm} \ell^{\mp}
\rightarrow \tilde{\chi}_1^0 \ell^+ \ell^-$ (solid line),
$\tilde{\chi}_2^0 \rightarrow \tilde{\chi}_1^0 \ell^+ \ell^-$ (dotted line),
$\tilde{\chi}_2^0 \rightarrow \tilde{\tau}_1^{\pm}
\tau^{\mp}\rightarrow \tau^+ \tau^- \tilde{\chi}_1^0
\rightarrow \ell^+ \nu_{\ell} \bar \nu_{\tau}\, \ell^-
\bar \nu_{\ell} \nu_{\tau} \tilde{\chi}_1^0\,$ (dashed-dotted line),  
$\tilde{\chi}_2^0 \rightarrow \tau^+ \tau^- \tilde{\chi}_1^0 \rightarrow 
\ell^+ \nu_{\ell} \bar \nu_{\tau} \,\ell^- 
\bar \nu_{\ell} \nu_{\tau} \tilde{\chi}_1^0$ (short dashed line) 
$\tilde{\chi}_2^0 \rightarrow  \tilde{\ell}_L^{\pm} \ell^{\mp}
\rightarrow \tilde{\chi}_1^0 \ell^+ \ell^-$ (long dashed line) 
with $\ell=e$ and $\mu$, for
$\tan \beta = 10, \, A_0=0, \, \mu > 0$.}
\end{figure}

\clearpage
\begin{figure}[htbp]
\centering
\resizebox{160mm}{180mm}
{\includegraphics{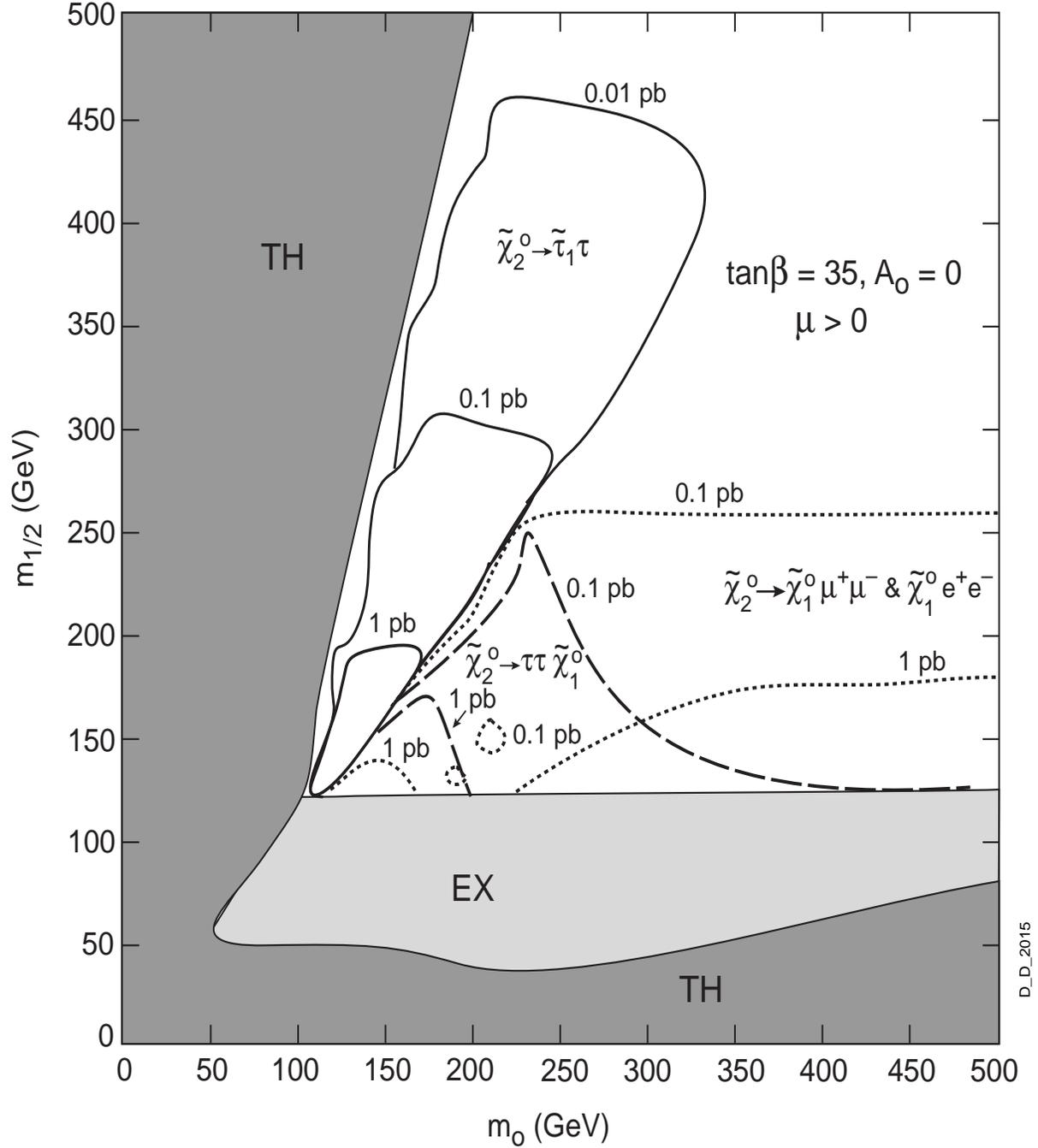}}
\caption{Contour lines for cross-section times branching ratios in the 
($m_0,m_{1/2}$) plane for indirect and associated $\tilde{\chi}_2^0$ production
followed by decays:
$\tilde{\chi}_2^0 \rightarrow  \tilde{\ell}_R^{\pm} \ell^{\mp}
\rightarrow \tilde{\chi}_1^0 \ell^+ \ell^-$ (solid line),
$\tilde{\chi}_2^0 \rightarrow \tilde{\chi}_1^0 \ell^+ \ell^-$ (dotted line),
$\tilde{\chi}_2^0 \rightarrow \tilde{\tau}_1^{\pm}
\tau^{\mp}\rightarrow \tau^+ \tau^- \tilde{\chi}_1^0
\rightarrow \ell^+ \nu_{\ell} \bar \nu_{\tau}\, \ell^-
\bar \nu_{\ell} \nu_{\tau} \tilde{\chi}_1^0\,$ (dashed-dotted line),  
$\tilde{\chi}_2^0 \rightarrow \tau^+ \tau^- \tilde{\chi}_1^0 \rightarrow 
\ell^+ \nu_{\ell} \bar \nu_{\tau} \,\ell^- 
\bar \nu_{\ell} \nu_{\tau} \tilde{\chi}_1^0$ (short dashed line) 
$\tilde{\chi}_2^0 \rightarrow  \tilde{\ell}_L^{\pm} \ell^{\mp}
\rightarrow \tilde{\chi}_1^0 \ell^+ \ell^-$ (long dashed line) 
with $\ell=e$ and $\mu$, for
$\tan \beta = 35, \, A_0=0, \, \mu > 0$.}
\end{figure}

\clearpage
\begin{figure}[htbp]
\centering
\resizebox{170mm}{190mm}
{\includegraphics{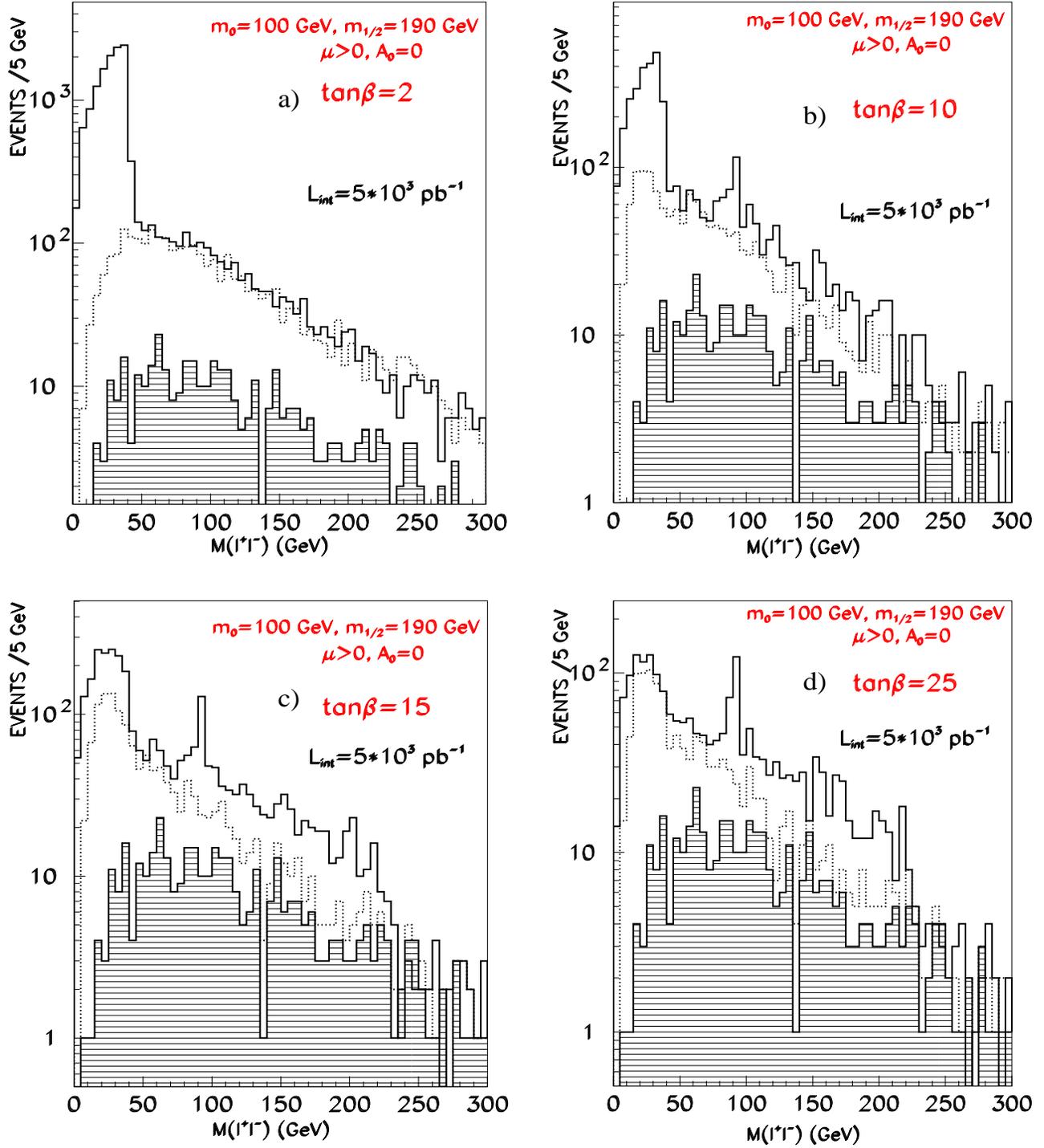}}
\caption{The invariant mass distributions of $e^+e^-$ and $\mu^+\mu^-$ 
(solid line) and  $e^{\pm}\mu^{\mp}$ (dotted line) 
lepton pairs at the point $m_0=100$ GeV, 
$m_{1/2}=190$ GeV,  $\mu>0$, $A_0=0$ with various  $\tan\beta=2,10,15,25$. 
SM background is also shown (hatched histogram).}
\end{figure}

\clearpage
\begin{figure}[htbp]
\centering
\resizebox{170mm}{190mm}
{\includegraphics{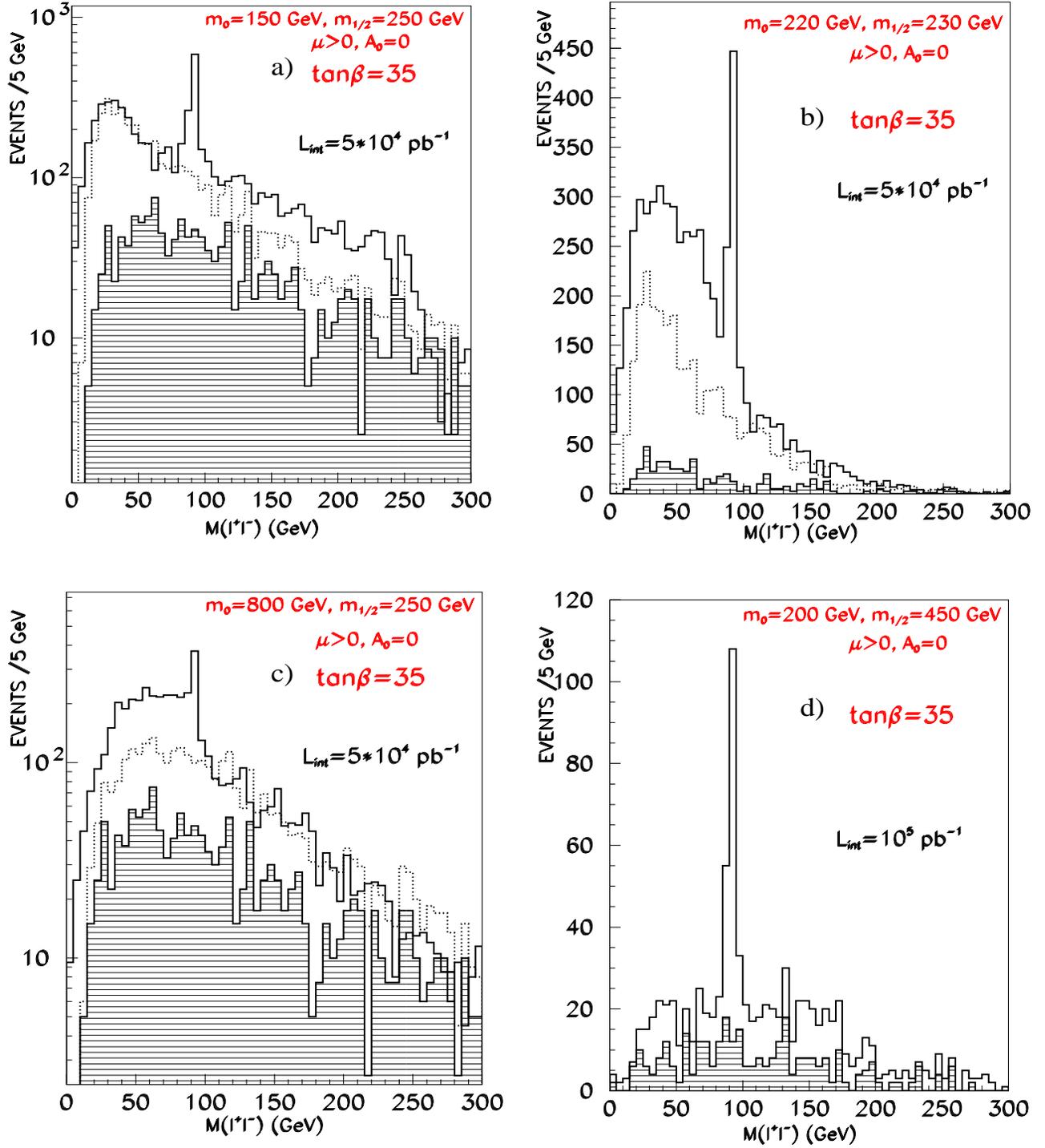}}
\caption{The invariant mass distributions of $e^+e^-$ and $\mu^+\mu^-$ 
(solid line) and  $e^{\pm}\mu^{\mp}$ (dotted line) lepton pairs at various
($m_0,m_{1/2}$) points with $\tan\beta=35$. 
SM background is also shown (hatched histogram).}
\end{figure}


\begin{thebibliography}{9}

\bibitem{Bartl} H. Baer, X. Tata, and J. Woodside, Phys. Rev. D \textbf{45},
142 (1992);\\
H. Baer, M. Bisset, X. Tata, and J. Woodside, {\em{ibid.}} \textbf{46},
303 (1992);\\
A. Bartl, W. Majerotto, B. M\"osslacher, and N. Oshimo, Z. Phys.
C \textbf{52}, 477 (1991);\\
C. Albajar et al., in {\em{Proceedings of the ECFA Large Hadron Collider
Workshop}}, Aachen, Germany, 1990, ed. by G. Jarlskog and D. Rein, CERN
Report No. 90-10, Geneva, Switzerland;\\
F. Pauss, {\em{ibid.}}\\
H. Baer et al., in {\em{Research Directions for the Decade, Proceedings of
the Summer Study}}, Snowmass, Colorado, 1990, ed. by E. Berger (World
Scientific, Singapore, 1992).

\bibitem{Dine} For more recent reviews see, e.g.: \\
M. Dine, hep--ph/9612389;\\
F.E. Paige, hep--ph/9801254;\\
S. Dawson, hep--ph/9712464;\\
J. Gunion, hep--ph/9801417;\\
R. Arnowitt and P. Nath, hep--ph/9708254;\\
S.P. Martin, hep--ph/9709356.

\bibitem{BaerPaige} F. Paige and S. Protopopescu, in {\em{Supercollider
Physics}}, p.41, ed. D. Soper (World Scientific, 1986);\\
H. Baer, F. Paige, S. Protopopescu, and X. Tata, in {\em{Proceedigns
of the Workshop on Physics at Current Accelerators and
Supercolliders}}, ed. J. Hewett, A. White, and D. Zeppenfeld (Argonne
National Laboratory, 1993).


\bibitem{Chen} H. Baer, C.-H. Chen, F. Paige, and X. Tata, Phys. Rev.
D \textbf{52}, 2756 (1995);\\
I. Hinchliffe, J. Womersley, LBNL--38997;\\
 H. Baer, X. Tata, and J. Woodside , Phys. Rev. D \textbf{53},
6241
(1996);\\

\bibitem{CMScollab} 
CMS presentation at the LHCC SUSY Workshop, CERN, Oct.
29--30, 1996, CMS Document 1996--149 (PH-SUSY);\\
S. Abdullin et al., CMS Note 1998/006, hep-ph/9806366.

\bibitem{Majer}  D. Denegri, W. Majerotto, L. Rurua, hep-ph/9711357,
Phys. Rev. D58, 095010 (1998).

\bibitem{Baer1}  H. Baer, C.-H. Chen, M. Drees, F. Paige, and X. Tata,
Phys. Rev. Lett. \textbf{79}, 986 (1997).

\bibitem{Bartl2} A. Bartl, W. Majerotto, W. Porod,
 Z. Phys. C \textbf{64}, 499 (1994);
                  err. Z. Phys. C \textbf{68}, 518 (1995).

\bibitem{Baer3} H. Baer, C.-H. Chen, M. Drees, F. Paige, and X. Tata,
 Phys. Rev. D \textbf{58}, 075008 (1998).

\bibitem{Baer4} H. Baer, C.-H. Chen, M. Drees, F. Paige, and X. Tata,
hep-ph/9809223.

\bibitem{Kakuto} K. Inoue, A. Kakuto, H. Komatsu, and S Takeshita, Prog.
Theor. Phys. \textbf{68}, 927 (1982);\\
M. Drees and S.P. Martin, hep--ph/9504324.


\bibitem{Barger} H. Baer, V. Barger, D. Karatas, and X. Tata, Phys. Rev. 
D \textbf{36}, 96 (1987);\\
R.M. Barnett, J.F. Gunion, and H.E. Haber, Phys. Rev. Lett. \textbf{60}, 
401 (1988); Phys. Rev. D \textbf{37}, 1892 (1988);\\
A. Bartl, W. Majerotto, B. M\"osslacher, N. Oshimo, and S. Stippel, 
Phys. Rev. D \textbf{43}, 2214 (1991);\\
A. Bartl, W. Majerotto, and W. Porod, Z. Phys. C \textbf{64}, 499 (1994).

\bibitem{Baer2} H. Baer, K. Hagiwara, X. Tata, Phys. Rev. D \textbf{35},
1598 (1987);\\
H. Baer, D.D. Karatas, X. Tata, Phys. Rev. D \textbf{42}, 2259 (1990)
[Fig. 6(a)];\\
H. Baer, C. Kao, X. Tata, Phys. Rev. D \textbf{48}, 5175 (1993);\\
H. Baer, C.-H. Chen, F. Paige, X. Tata, Phys. Rev. D \textbf{50},
4508 (1994).
 
 
\bibitem{Paige} F. Paige, Determining SUSY particle masses at LHC, Proc.
of the 1996 DPF/DPB Summer Study on High-Energy Physics `New
Directions for High-Energy Physics', Snowmass, Colorado, 1996,
p.710;\\
A. Bartl et al., Supersymmetry at LHC, {\em{ibid.}}, p.693;\\
J. Amundson et al., Report of the Supersymmetry Theory Subgroup,
{\em{ibid.}}, p.655; \\
I. Hinchliffe, F.E. Paige, M.D. Shapiro,
             J. S\"{o}derqvist, W. Yao, hep-ph/9610544.
 
\bibitem{Sjostrand} T. Sj\"{o}strand,
  {\em Comp. Phys. Com.} \textbf{39}, 347 (1986);
 
  T. Sj\"{o}strand and M. Bengtsson,
  {\em Comp. Phys. Com.} \textbf{43}, 367 (1987);
 
  H.U. Bengtsson and T. Sj\"{o}strand,
  {\em Comp. Phys. Com.} \textbf{46}, 43 (1987);
 
  T. Sj\"{o}strand, CERN-TH.7112/93.

\bibitem{Khanov} S. Abdullin, A. Khanov and N. Stepanov, CMS TN/94-180.
\end{thebibliography}
\end{document}